\documentclass[a4paper,conference]{IEEEtran}
\ifCLASSINFOpdf
\else
\fi
\usepackage[pdftex]{graphicx}
\usepackage{cite}
\usepackage[caption=false,font=footnotesize]{subfig}

\newcommand{\wfig}[1]{Fig.\ref{fig:#1}}
\newcommand{\wfigure}[1]{Figure \ref{fig:#1}}

\newcommand{\wtable}[1]{Table \ref{tab:#1}}

\begin{document}
%
\title{Extension of JPEG XS \\for Two-Layer Lossless Coding}

\author{\IEEEauthorblockN{Hiroyuki KOBAYASHI}
\IEEEauthorblockA{
 Tokyo Metropolitan College  of Industrial Technology,\\
 Email: hkob@metro-cit.ac.jp}
\and
\IEEEauthorblockN{Hitoshi KIYA}
\IEEEauthorblockA{
 Tokyo Metropolitan University\\
Email: kiya@tmu.ac.jp}
}


%


\maketitle



%
\IEEEpeerreviewmaketitle

\section*{Abstract}

A two-layer lossless image coding method compatible with JPEG XS is proposed.
JPEG XS is a new international standard for still image coding that has the characteristics of very low latency and very low complexity.
However, it does not support lossless coding, although it can achieve visual lossless coding.
The proposed method has a two-layer structure similar to JPEG XT, which consists of JPEG XS coding and a lossless coding method.
As a result, it enables us to losslessly restore original images, while maintaining compatibility with JPEG XS. 

\begin{IEEEkeywords}
JPEG XS, lossless coding, two-layer coding
\end{IEEEkeywords}

\section{Introduction}

JPEG XS was standardized as a new still image coding method\cite{JPEG_XS}.
This standard is also expected to be applied to videos, and enables us to compress images with low latency and low complexity.
The coding aims to realize visual lossless quality, so it is not guaranteed to achieve lossless coding.

There are many applications that require lossless coding, such as medical images, and master data of the cinema and TV programs.
In addition, lossless coding allows us to combine coding with other
technologies  such as data hiding and encryption\cite{Fujiyoshi2006,KURIHARA2017}.
Although numerous encodings such as JPEG-LS\cite{JPEG-LS}, JPEG 2000\cite{JPEG2000}, and JPEG XR\cite{JPEGXR} have been standardized for supporting lossless coding, conventional encoding methods have not considered the features of low latency and low cost that JPEG XS has.

Two-layer codings have been researched as a method of combining the characteristics of several codings
\cite{6411962,6288143,SehoonYea,Okuda2007,6637869,JPEGXT,8456254,KOBAYASHI2019,Kobayashi2019a,KobayashiGCCE2019}．
JPEG XT is a two-layer coding that uses JPEG as the first base layer in consideration of the compatibility with past JPEG decoders.
The second extension layer holds the residual image between the original image and the base layer decoded image.
In addition, JPEG XT Part 8\cite{JPEGXTpart8}, which is the the extension of JPEG XT, encodes the difference information losslessly.
As a result, losslessness of the bitstream can be realized.

Because of such a situation, we propose extending JPEG XS for supporting lossless coding, while maintaining the features of JPEG XS.
The extended coding has a two-layer structure, where the first layer, called base layer corresponds to the JPEG XS coding, and the second one, called extension layer is used for compressing residual data between an original image and the decoded image from the base layer.
This two-layer structure has been inspired by JPEG XT Part 8\cite{JPEGXTpart8} and its extension\cite{8456254}.
The proposed coding is compatible with JPEG XS.
In an experiment, the proposed coding is demonstrated not only to have compatibility with JPEG XS, but also to achieve lossless coding.


\section{JPEG XS}

JPEG XS is a new standard for still image coding\cite{JPEG_XS}.
This standard is intended for low latency and low complexity encoding, and is expected to be applied to moving picture coding in which each frame is regarded as an independent still image.
JPEG XS aims at encoding at a compression ratio of about 1/2 to 1/10 while maintaining visual lossless image quality, not improving the compression ratio for low bitrates.

The JPEG XS encoding uses the wavelet transform that is also used in JPEG 2000.
However, the processing in the vertical direction is suppressed to a few lines, thereby realizing low latency and low complexity in encoding and decoding.
Furthermore, since there is no frame buffer for the entire image, it can be implemented at low cost.

JPEG XS supports visual lossless coding, but does not support lossless one.
\wtable{losslessBitrate} shows examples of lossless coding for JPEG2000, JPEG LS, JPEG XR and JPEG XS.
In the table, image `lena' was losslessly compressed by JPEG XS, but the image `Moss' was not done. In contrast, other compression methods losslessly compressed all images.

\begin{table}[tb]
\caption{Bitrates of lossless coding}
\label{tab:losslessBitrate}
\begin{tabular}{ccccc}
Image & JPEG XS & JPEG 2000 & JPEG LS & JPEG XR \\\hline
lena(24[bits]) & 22.0 & 13.62 & 13.57 &	14.10\\
Moss(30[bits]) & Not realized & 19.19 & 20.86 & 23.74\\
Moss(36[bits]) & Not realized & 26.30 & 25.55 & 26.67\\
\end{tabular}
\end{table}

\section{Proposed Method}

\begin{figure}[tbp]
 \centering
 \includegraphics[width=0.45\textwidth]{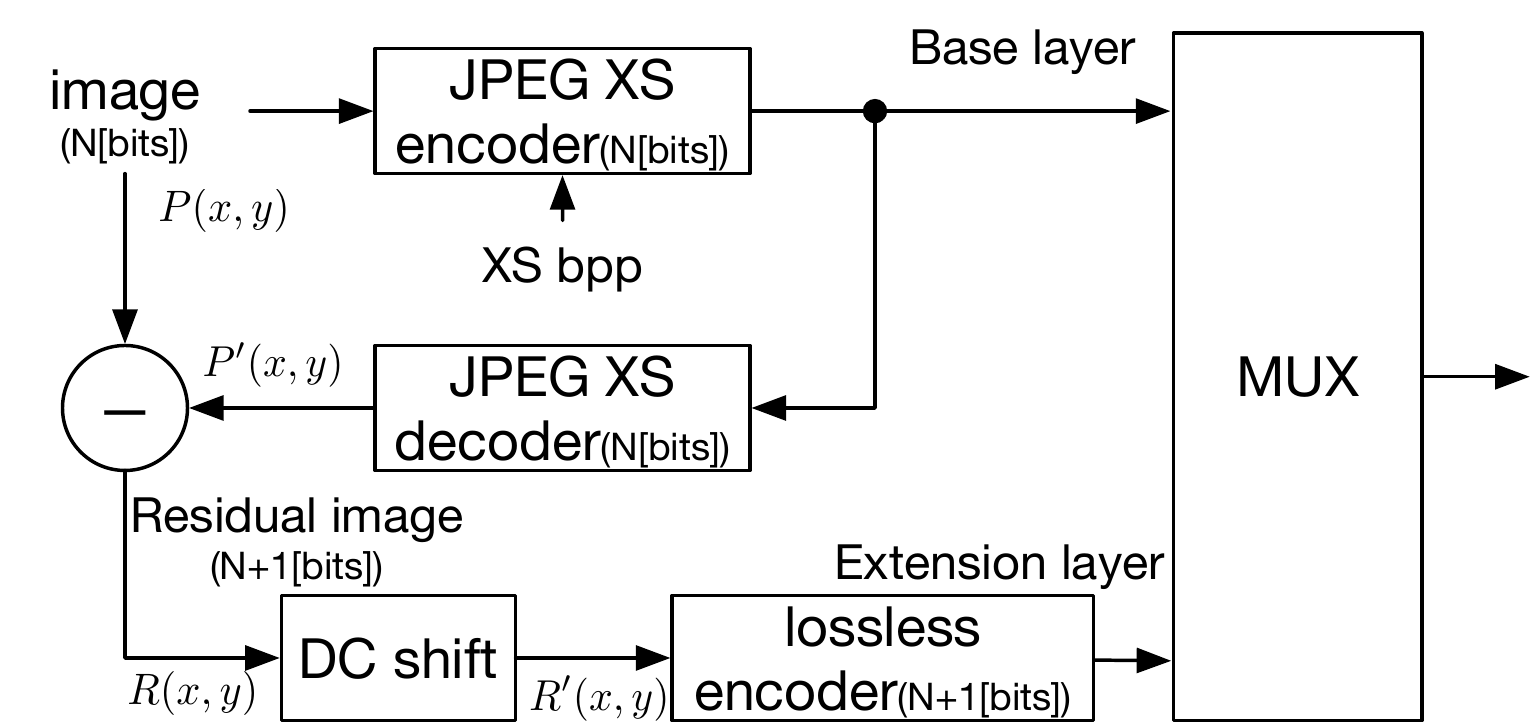} \\
 (a) encoder\\
 \includegraphics[width=0.45\textwidth]{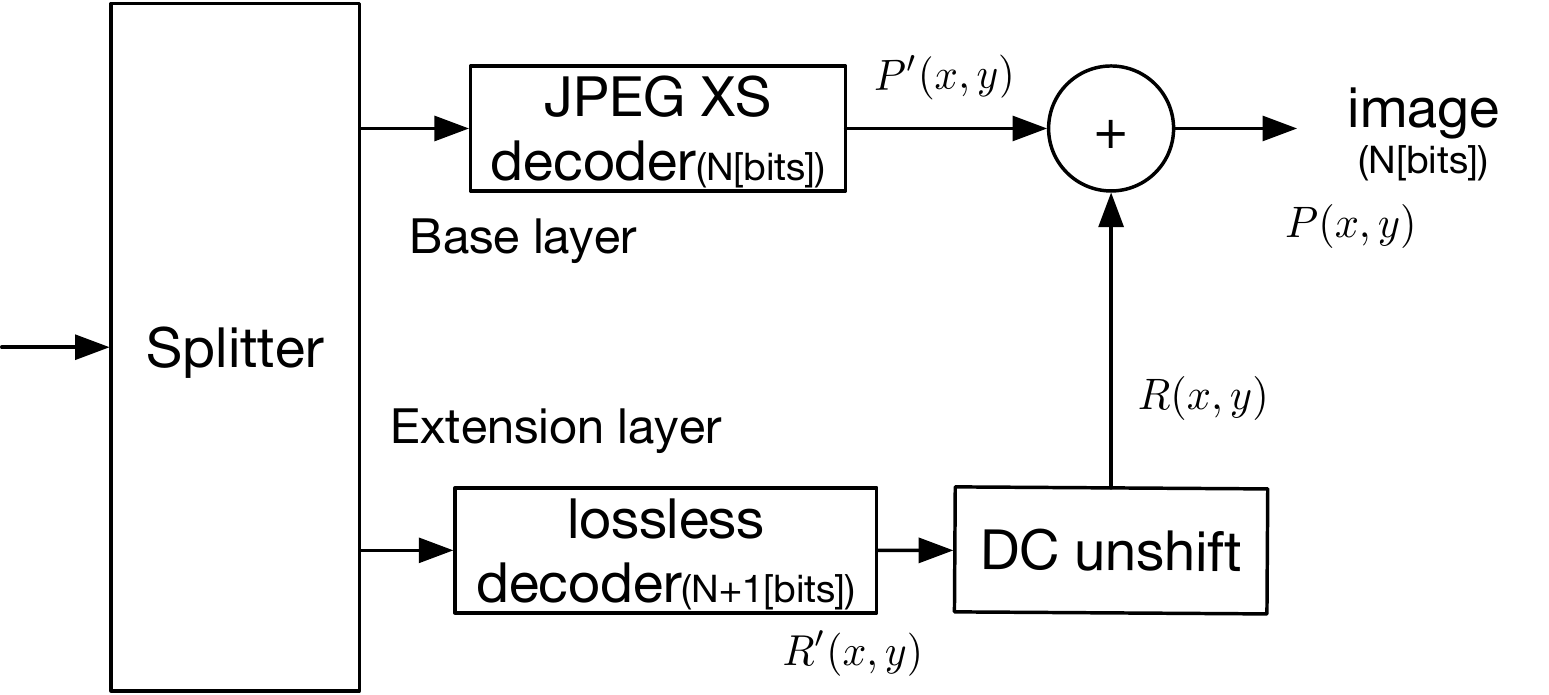} \\
 (b) decoder\\
 \caption{Block diagram of proposed method}
 \label{fig:proposed}
\end{figure}

As mentioned above, JPEG XS can not encode images losslessly.
Therefore, we consider two-layer coding that consists of a base layer and an extension layer.
\wfigure{proposed}(a) shows the encoder structure of the proposed lossless two-layer coding for $N$-bit-images.
The coding-path for generating the base layer is backward compatible with JPEG XS.
For the extension layer, the residual image $R(x, y)$ is generated by calculating the difference between decoded base layer image $P'(x, y)$ and the original image $P(x, y)$ as
\begin{equation}
R(x, y) = P(x, y) - P'(x, y).
\end{equation}
The residual data $R(x,y)$ include negative values, but lossless image compression methods do not support image with negative pixel values in general.
Therefore, all pixel values in $R(x,y)$ are shifted by the DC shift operation as
\begin{equation}
R'(x, y) = R(x, y) + 2^N-1,
\end{equation}
where  $R'(x, y)$ is expressed by using $N+1$ bits.
After the DC shifting operation, $R'(x, y)$ is encoded by using a lossless encoder such as JPEG-LS, JPEG 2000, and JPEG XR.

\wfigure{proposed}(b) shows the decoder structure of the proposed method.
Bitstreams of the base layer are decoded by JPEG XS and ones of the extension layer are decoded by a lossless decoder.
A residual image is reconstructed from the decoded image by using the DC inverse-shift operation, and is added to an image from the base layer.
Since the residual image is decoded losslessly, the final output image is also perfectly reconstructed. 

\section{Experimental results}
\begin{figure}[tb]
  \begin{tabular}{cc}
  \includegraphics[width=0.22\textwidth]{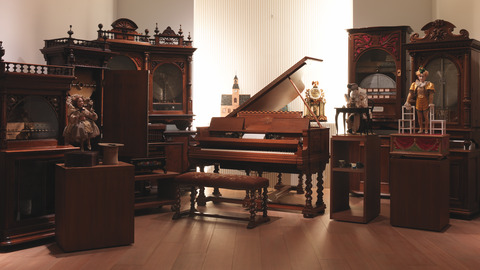} &
  \includegraphics[width=0.22\textwidth]{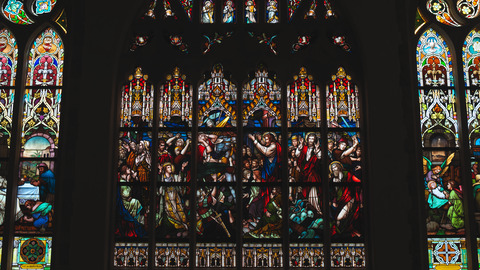} \\
  (a) MusicBox &
  (b) StainedGlass \\
  \includegraphics[width=0.22\textwidth]{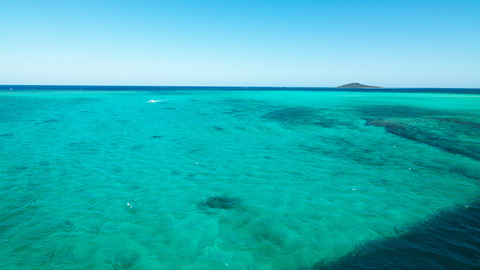} &
  \includegraphics[width=0.22\textwidth]{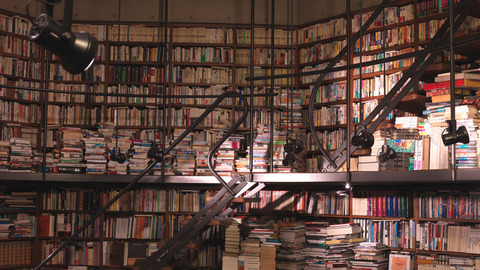} \\
  (c) Sea &
  (d) Books \\
  \includegraphics[width=0.22\textwidth]{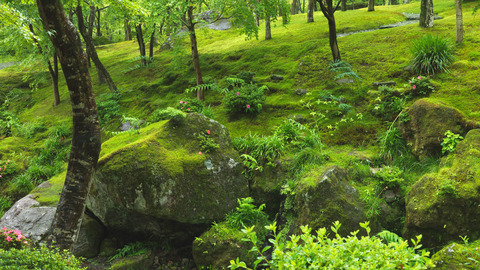} &
  \includegraphics[width=0.22\textwidth]{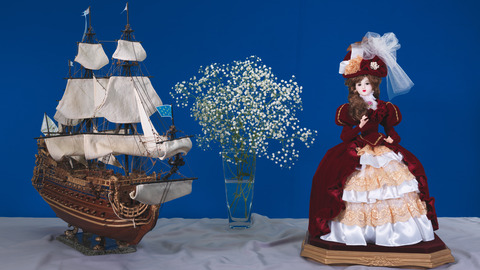} \\
  (e) Moss &
  (f) ChromaKey \\
  \end{tabular}
  \caption{Test images}
  \label{fig:testImages}
\end{figure}

\begin{figure}[tb]
  \includegraphics[width=0.44\textwidth]{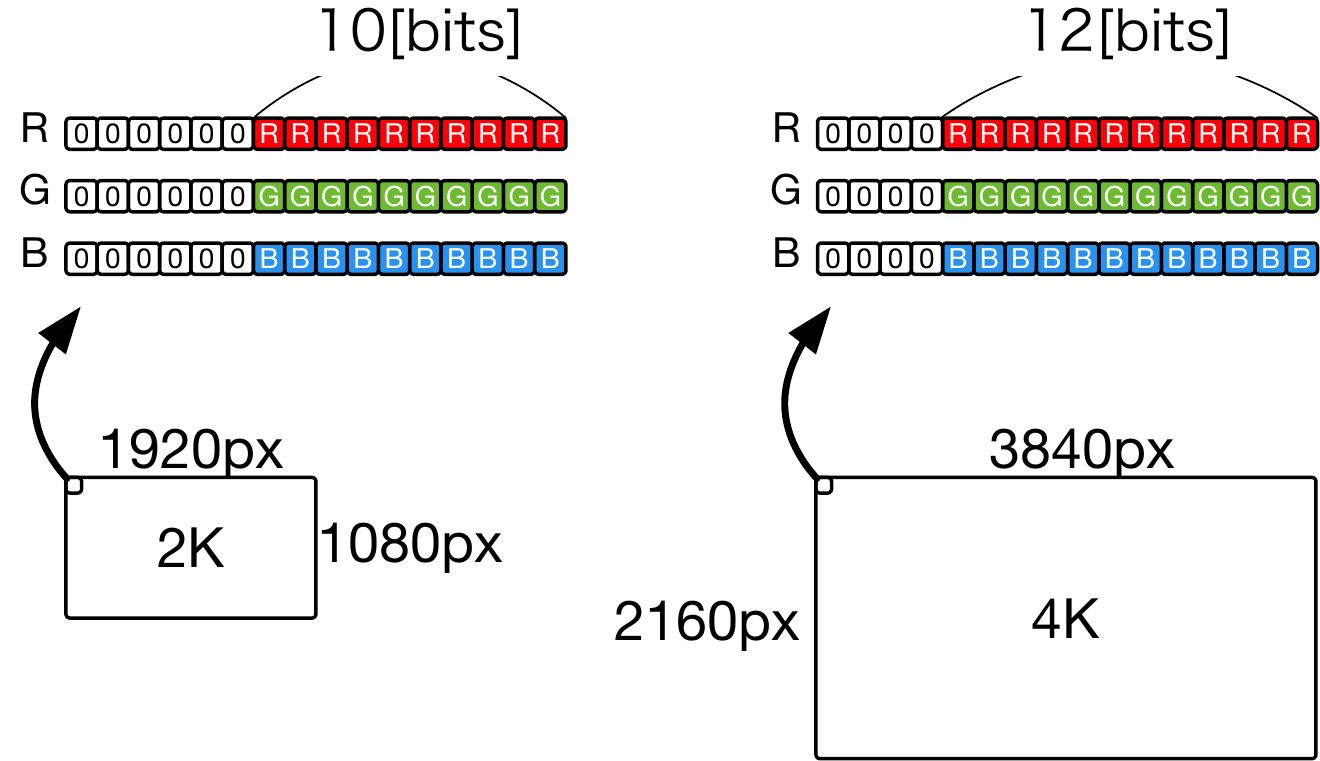}
  \caption{File formats}
  \label{fig:fileFormats}
\end{figure}

The compression performance of the proposed method was compared with two one-layer lossless codings: JPEG-LS and JPEG 2000.
In the experiment, the reference softwares provided by the JPEG committee were used.
Six 2K images with a depth of 30 bits and six 4K images with a depth of 36 bits provided from the Institute of Image Information and Television Engineers (ITE)\cite{TestChartHDR_e} were used in this experiment.
\wfigure{testImages} shows the six thumbnail images of 2K and 4K images and \wfig{fileFormats} shows the file formats for 2K and 4K images.
The 6 MSB bits in the 2K images and 4 MSB bits in the 4K images are filled with zero bits. 
In this paper, the six images are classified into two sets for convinience.
Set 1 has `MusicBox', `StainedGlass' and `Sea', and Set 2 has `Books', `Moss', and `ChromaKey'.

\subsection{JPEG XS image quality}
\begin{figure}[tbp]
 \centering
 \begin{tabular}{cc}
 \includegraphics[width=0.22\textwidth]{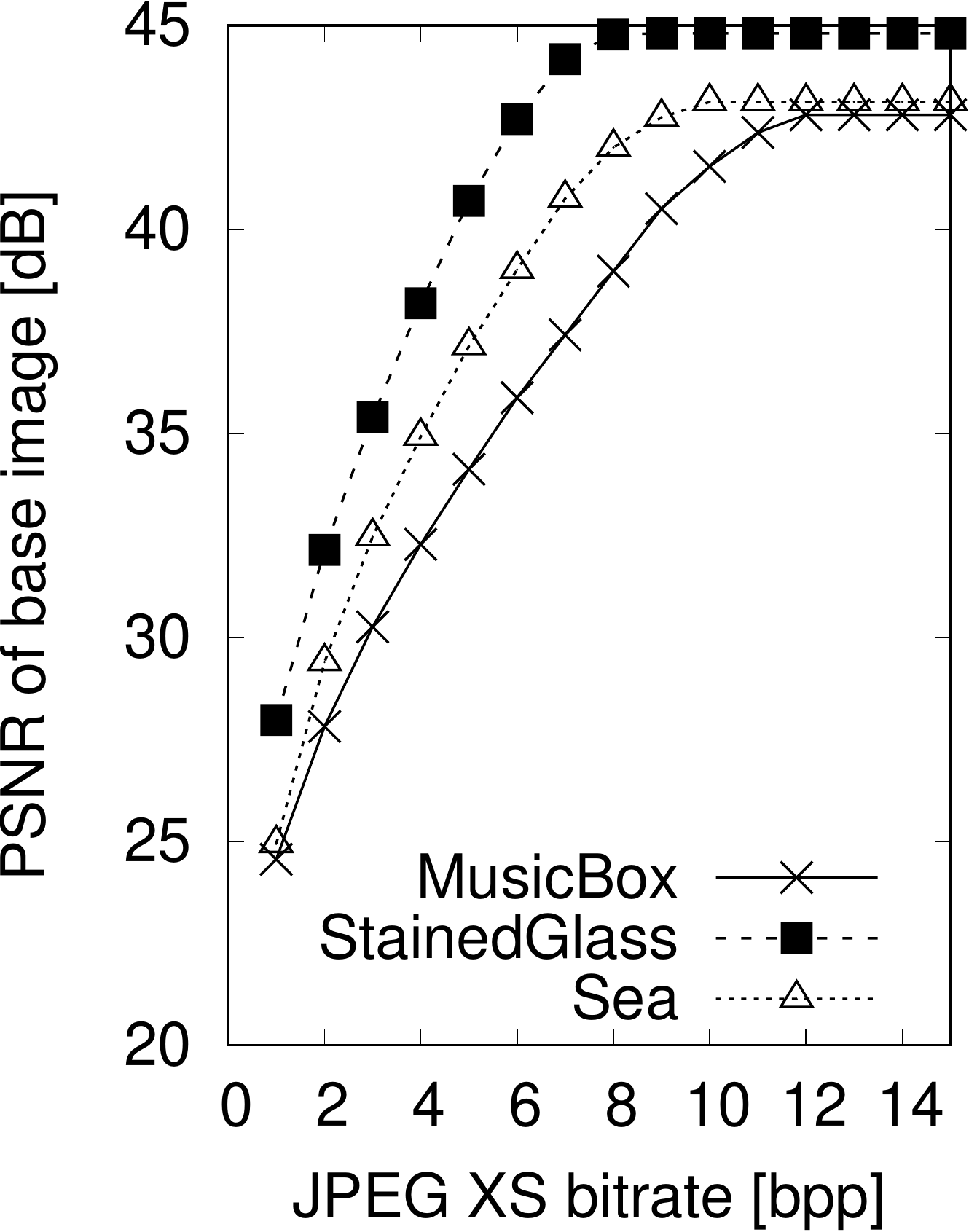} &
 \includegraphics[width=0.22\textwidth]{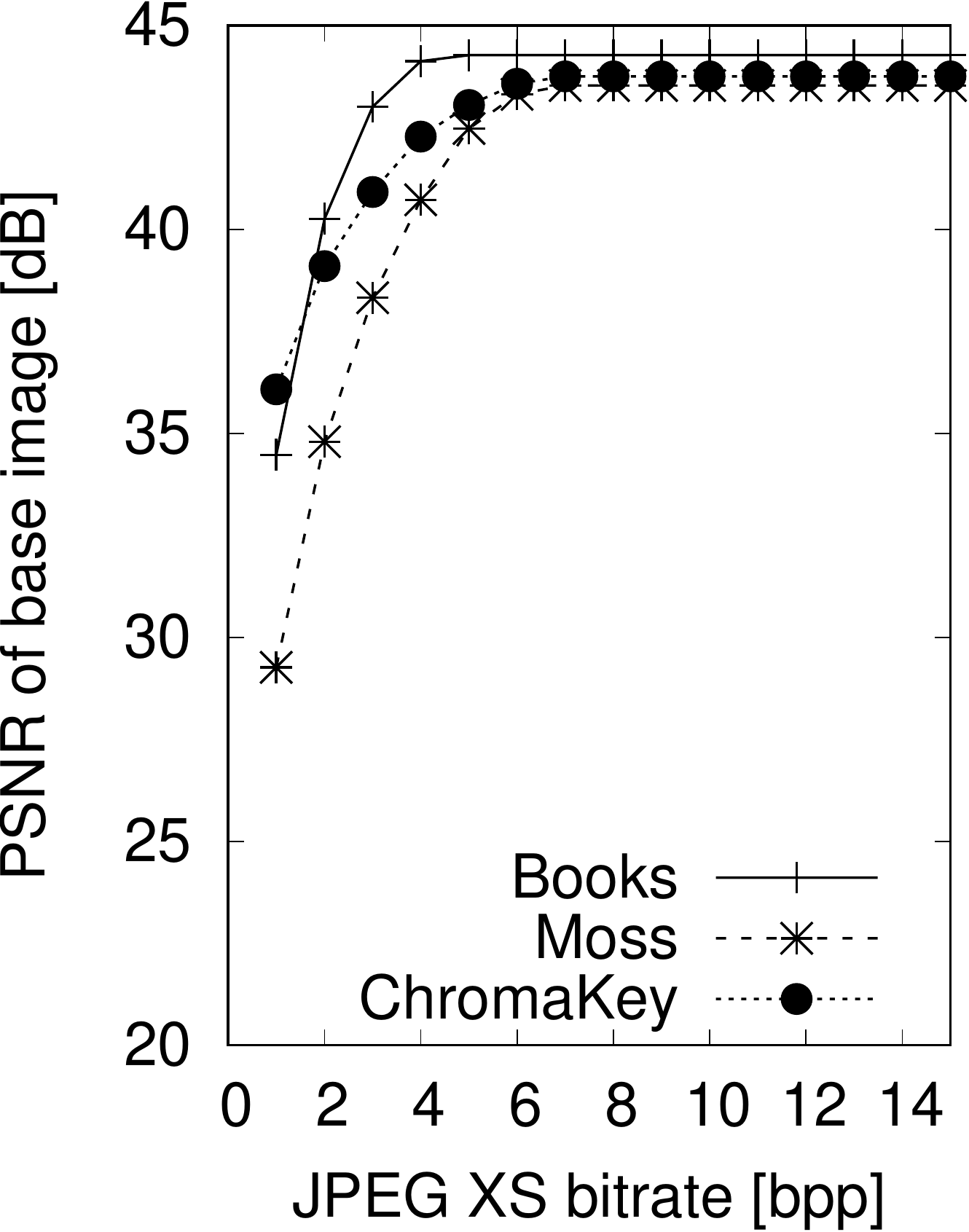}\\
 (a) 2K-images (Set 1) & (b) 2K-images (Set 2) \\
 \includegraphics[width=0.22\textwidth]{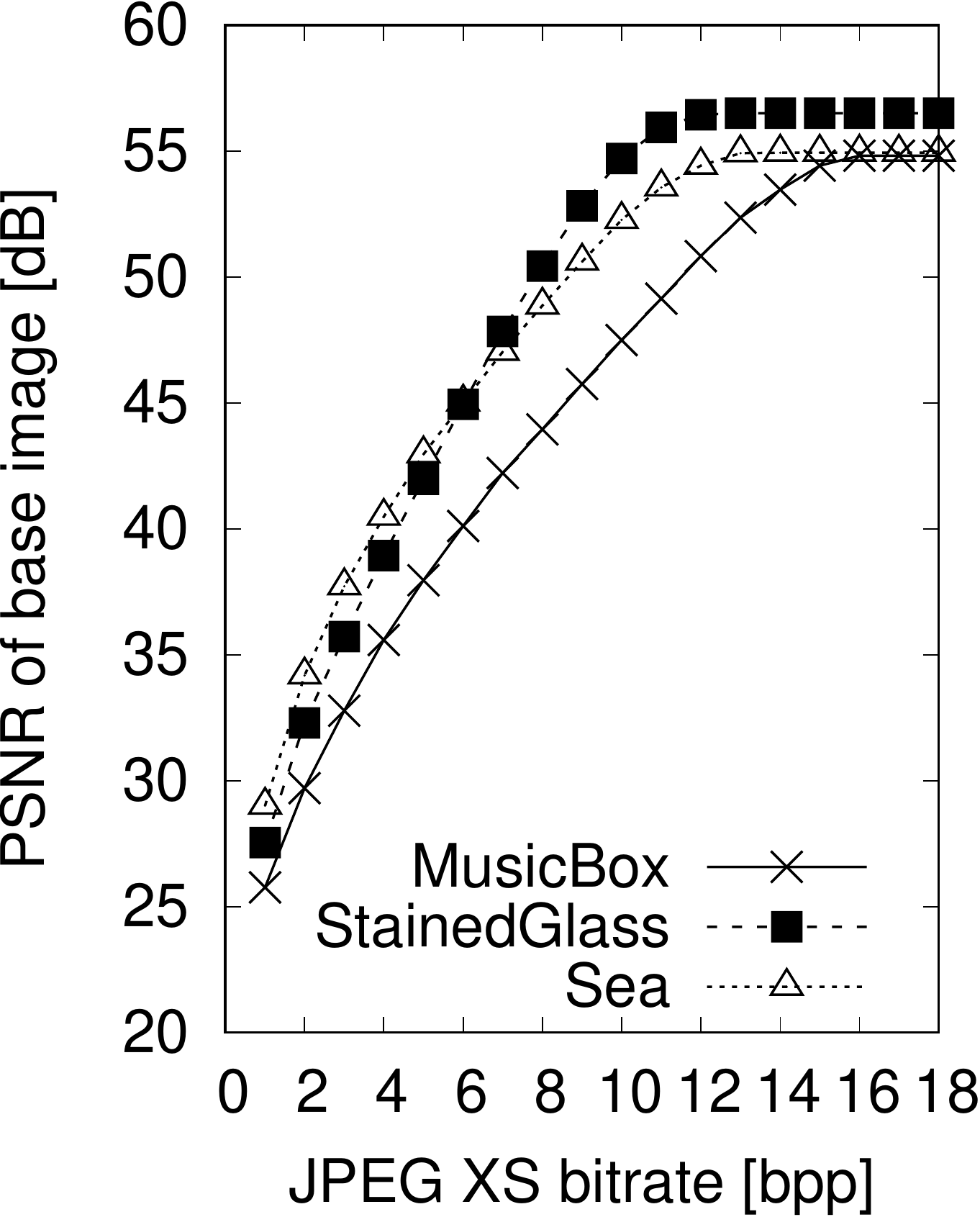} &
 \includegraphics[width=0.22\textwidth]{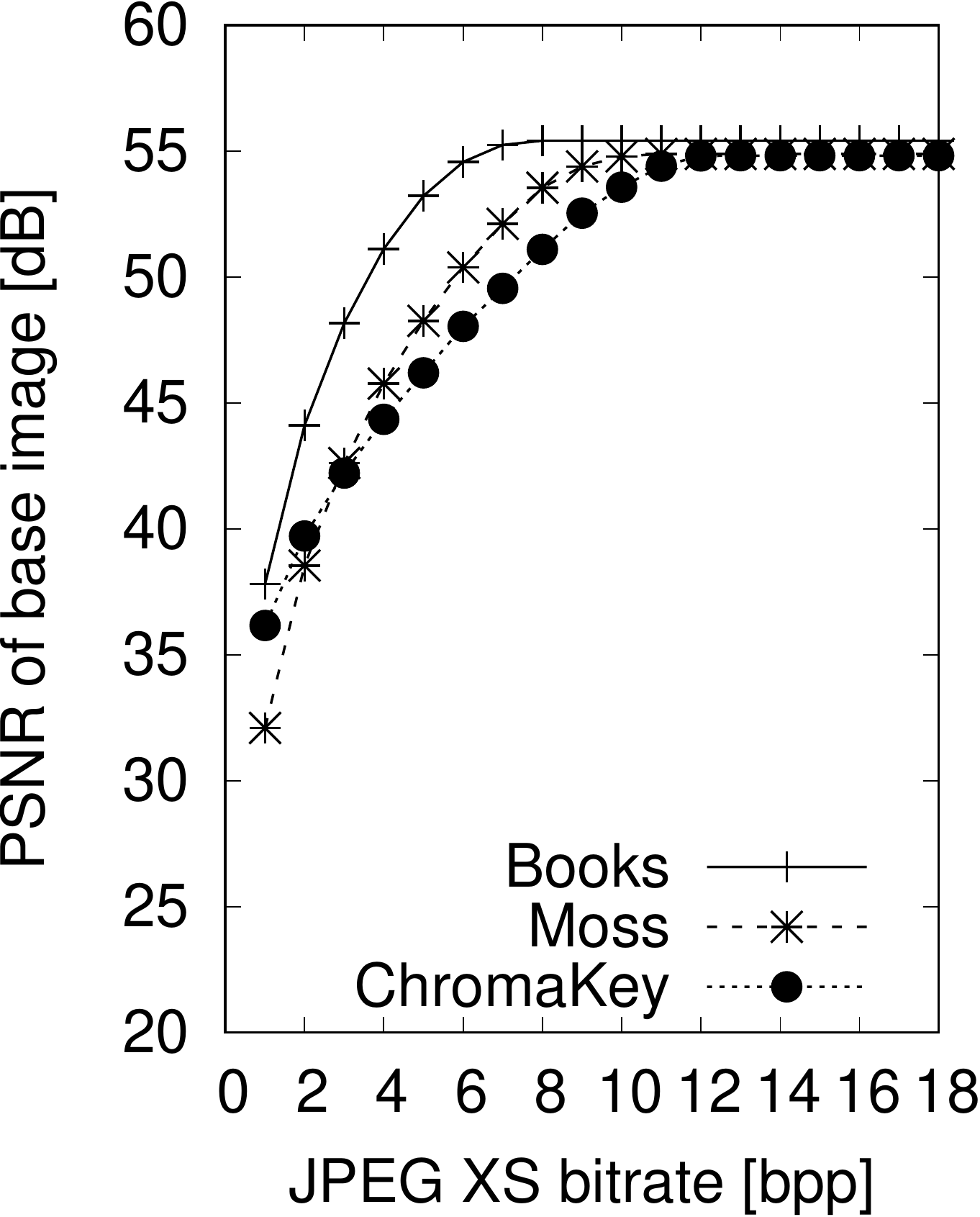}\\
 (c) 4K-images (Set 1) & (d) 4K-images (Set 2) \\
 \end{tabular}
 \caption{Rate distortion curves of JPEG XS}
 \label{fig:PSNR}
\end{figure}

At first, the quality of base-layer images, which are produced from base layer bitstreams by using the JPEG XS decoder, is addressed.
\wfigure{PSNR} shows rate distortion curves of reconstructed base-layer images.

For all images, PSNR values saturated at a certain PSNR value.
In other words, all images were not compressed losslessly, even when bitrate values increased.
In particular, image `Books' saturated at 4[bpp] in 2K images and at 6[bpp] in 4K images.

\subsection{Total bitrates of two-layer coding with JPEG 2000}

\begin{figure}[tbp]
 \centering
 \begin{tabular}{cc}
  \includegraphics[width=0.22\textwidth]{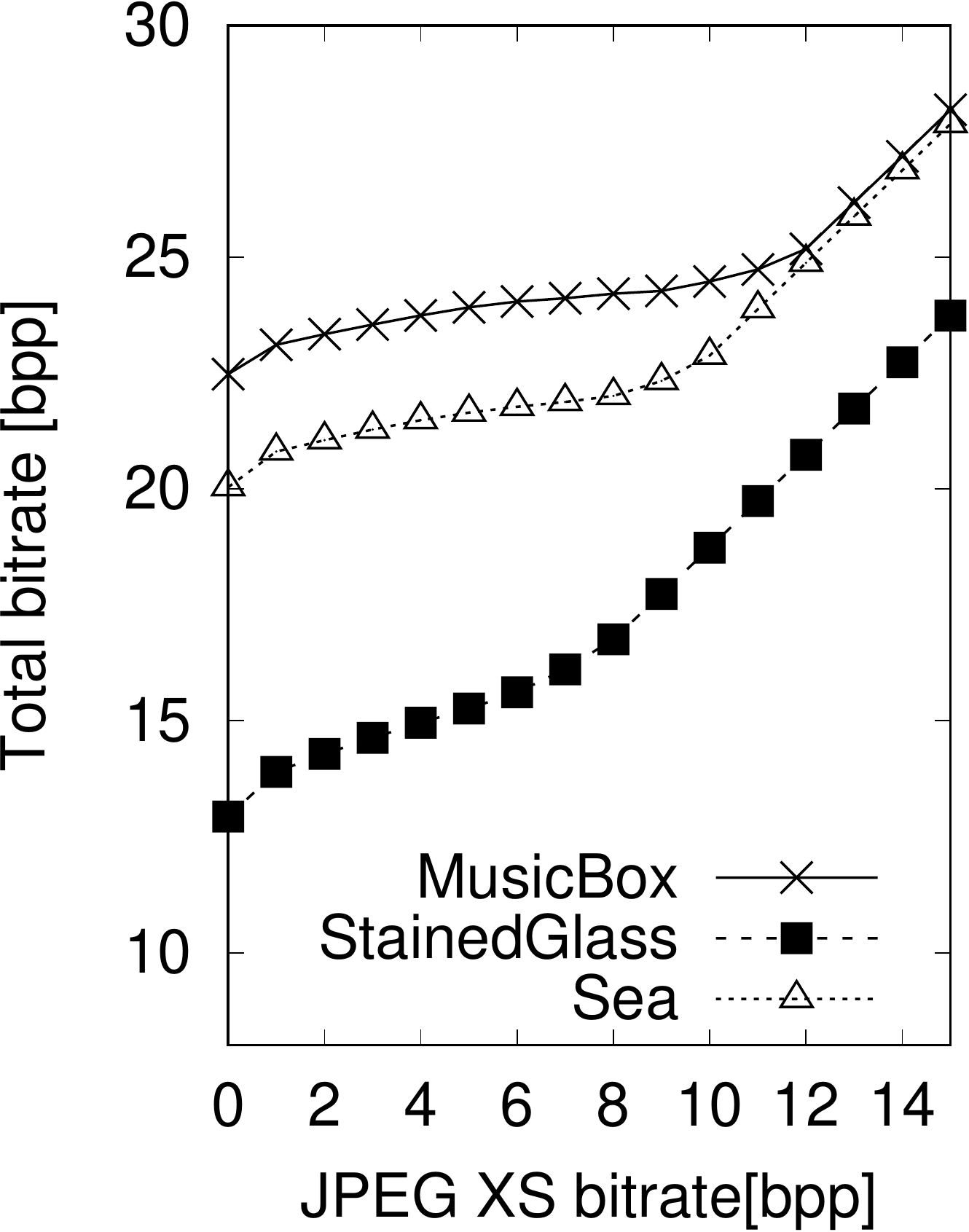} &
  \includegraphics[width=0.22\textwidth]{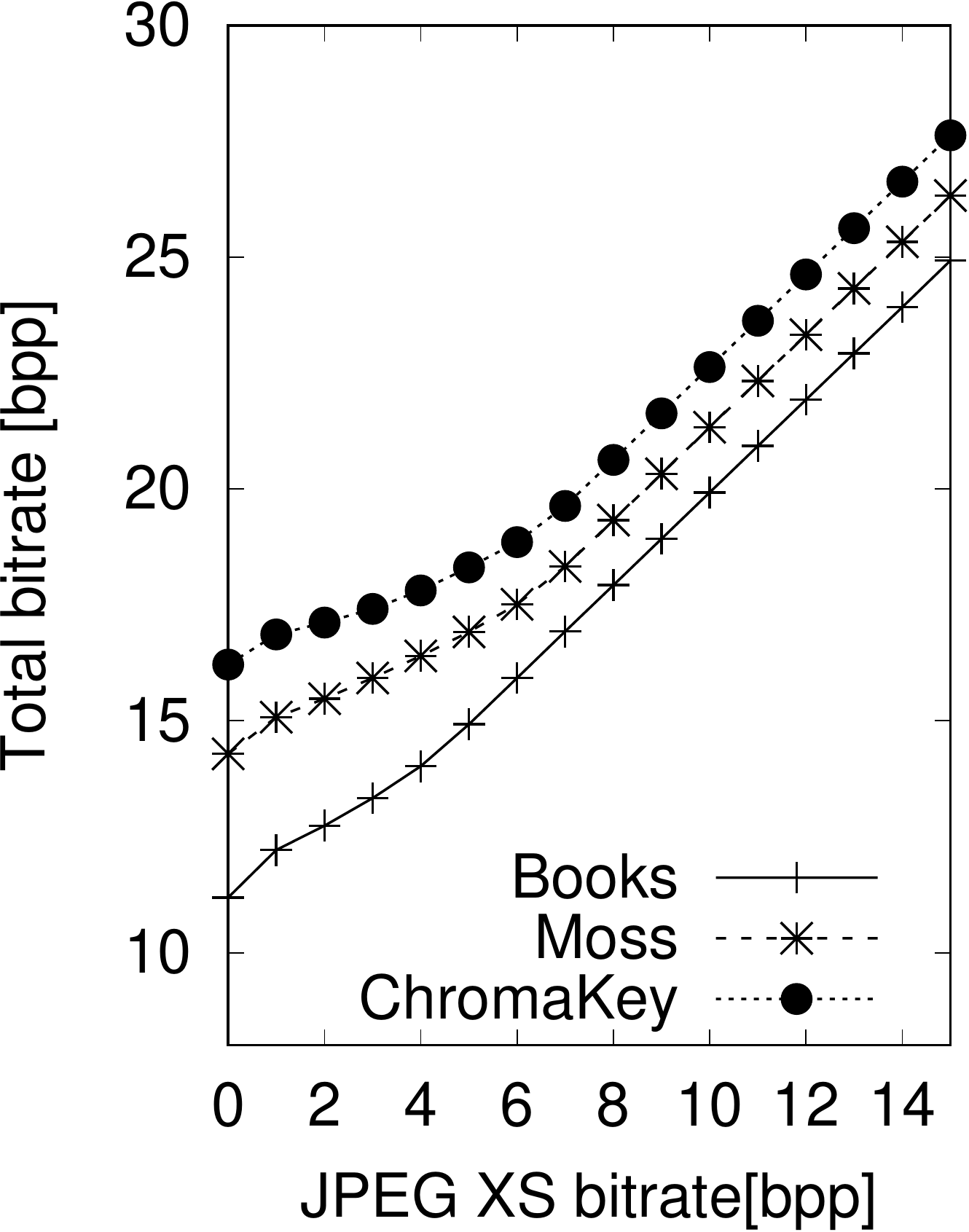} \\
  (a) 2K-images (Set 1) &
  (b) 2K-images (Set 2) \\ 
  \includegraphics[width=0.22\textwidth]{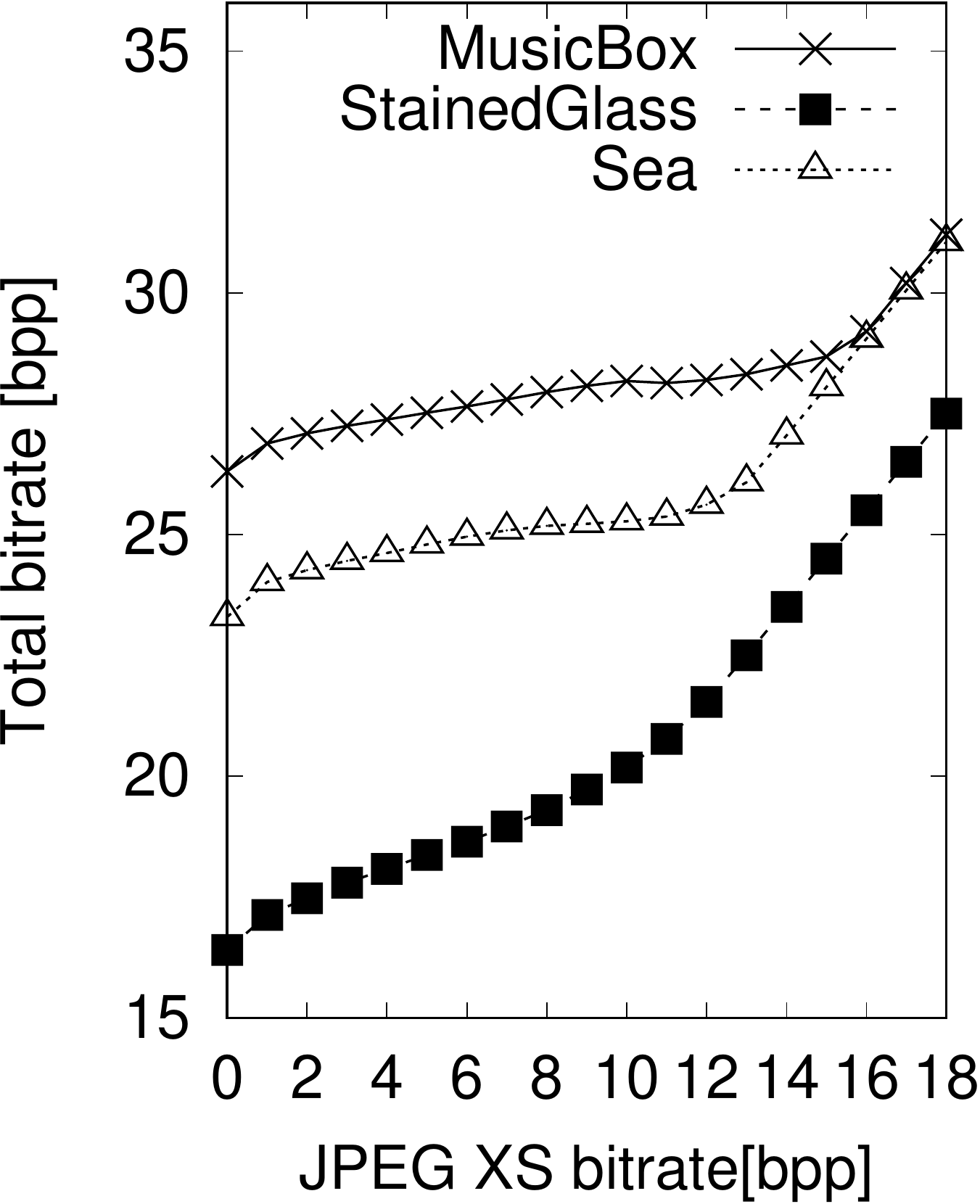} &
  \includegraphics[width=0.22\textwidth]{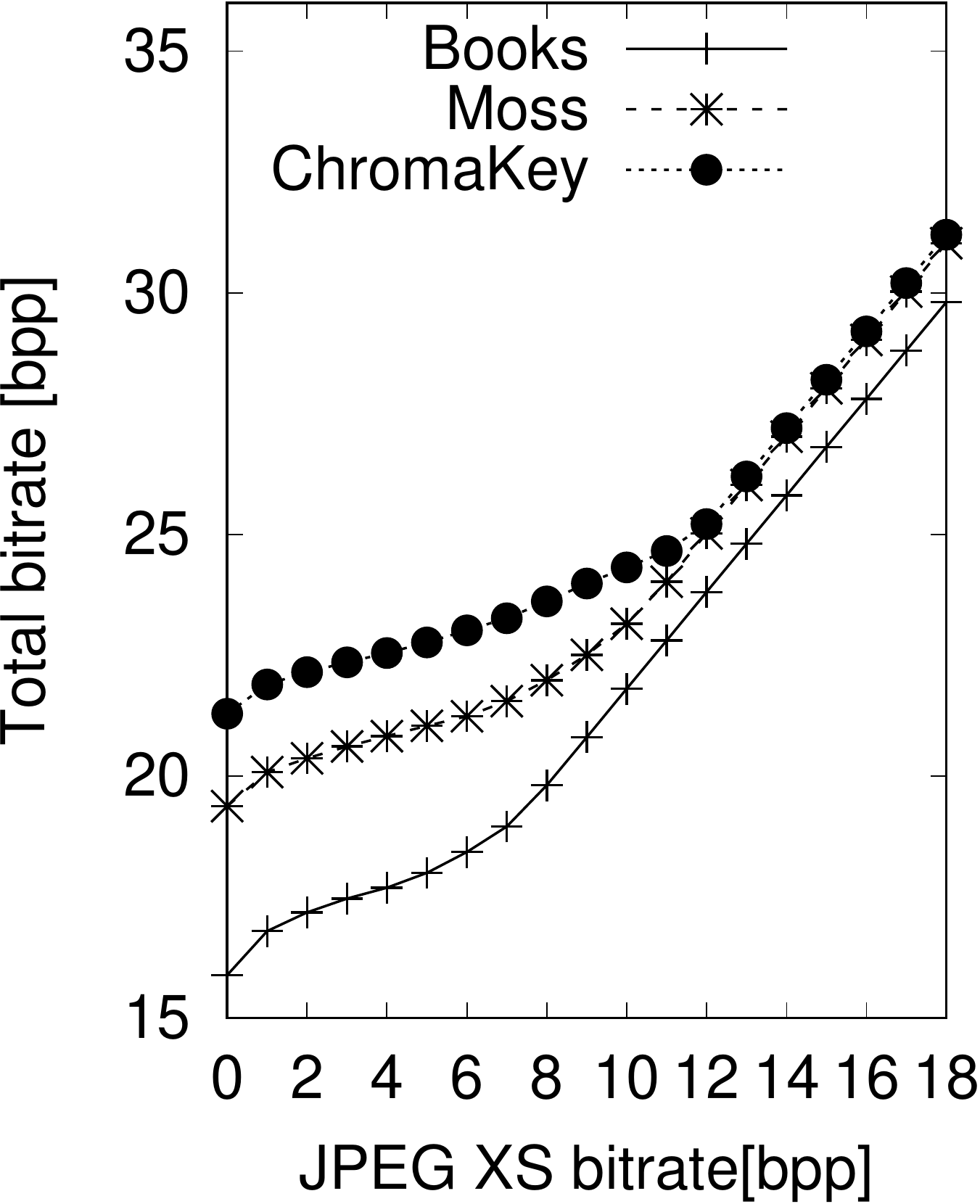} \\
  (c) 4K-images (Set 1) &
  (d) 4K-images (Set 2) \\
 \end{tabular}
 \caption{Total bitrates of proposed coding with JPEG 2000}
 \label{fig:compBitrate}
\end{figure}

\wfigure{compBitrate} shows total bitrates of the proposed two-layer coding under various bitrates of JPEG XS, where zero value in the horizontal axis corresponds to lossless coding without the base layer. From the results, the proposed coding was demonstrated to achieve lossless coding under all conditions. Besides, the total bitrate values increase, compared with those of using only JPEG 2000 without the two-layers structure.


\subsection{Comparison of lossless coders}
The proposed two-layer coding allows us to use an arbitrary lossless encoder.
Compared with JPEG 2000, JPEG-LS has low-delay and low-complexity.
The two-layer coding with JPEG-LS was compared with that with JPEG 2000 in terms of compression performance. 

\begin{figure}[tbp]
 \centering
 \begin{tabular}{c}
  \includegraphics[width=0.44\textwidth]{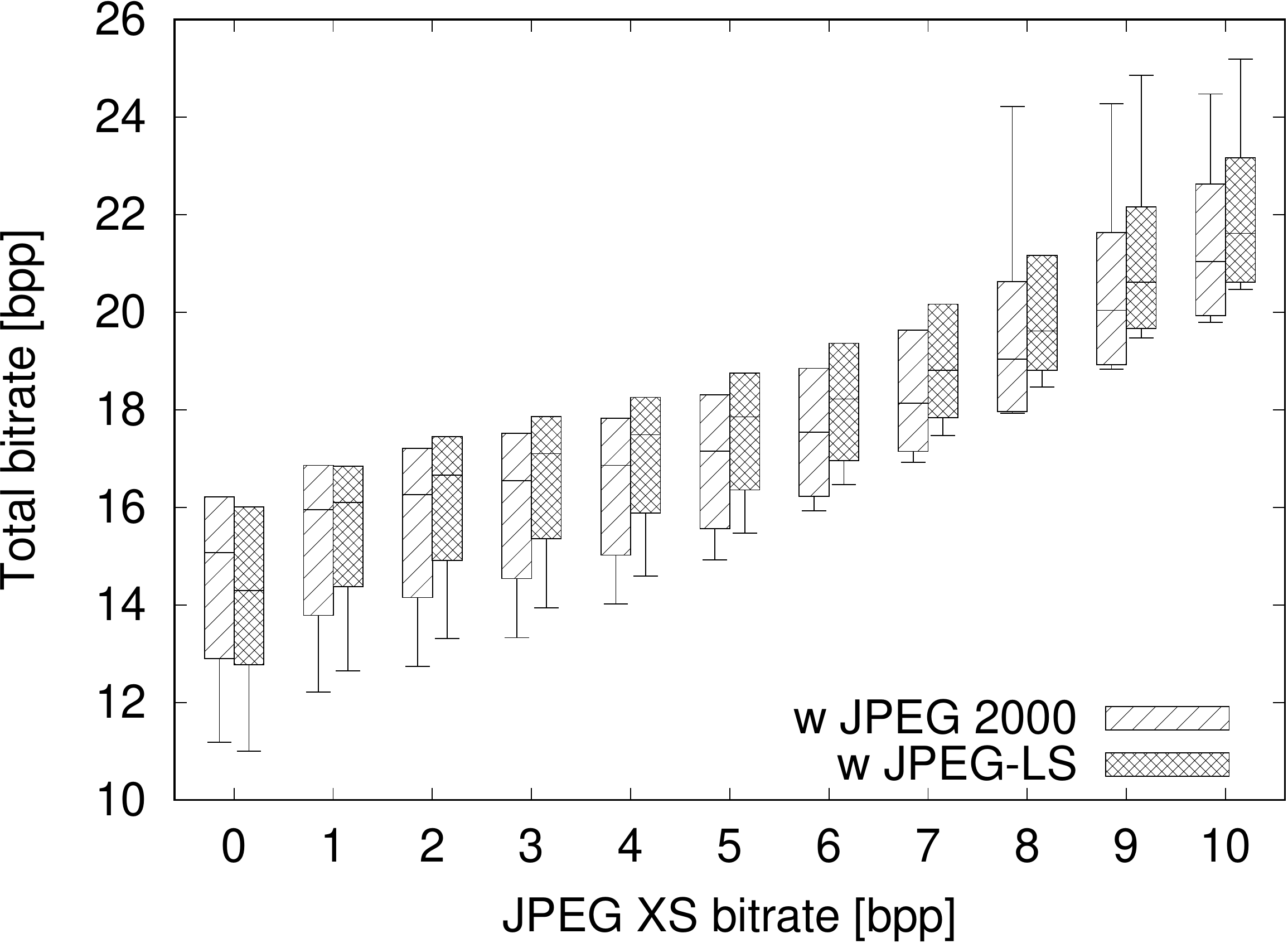} \\
  (a) 2K-images \\
  \includegraphics[width=0.44\textwidth]{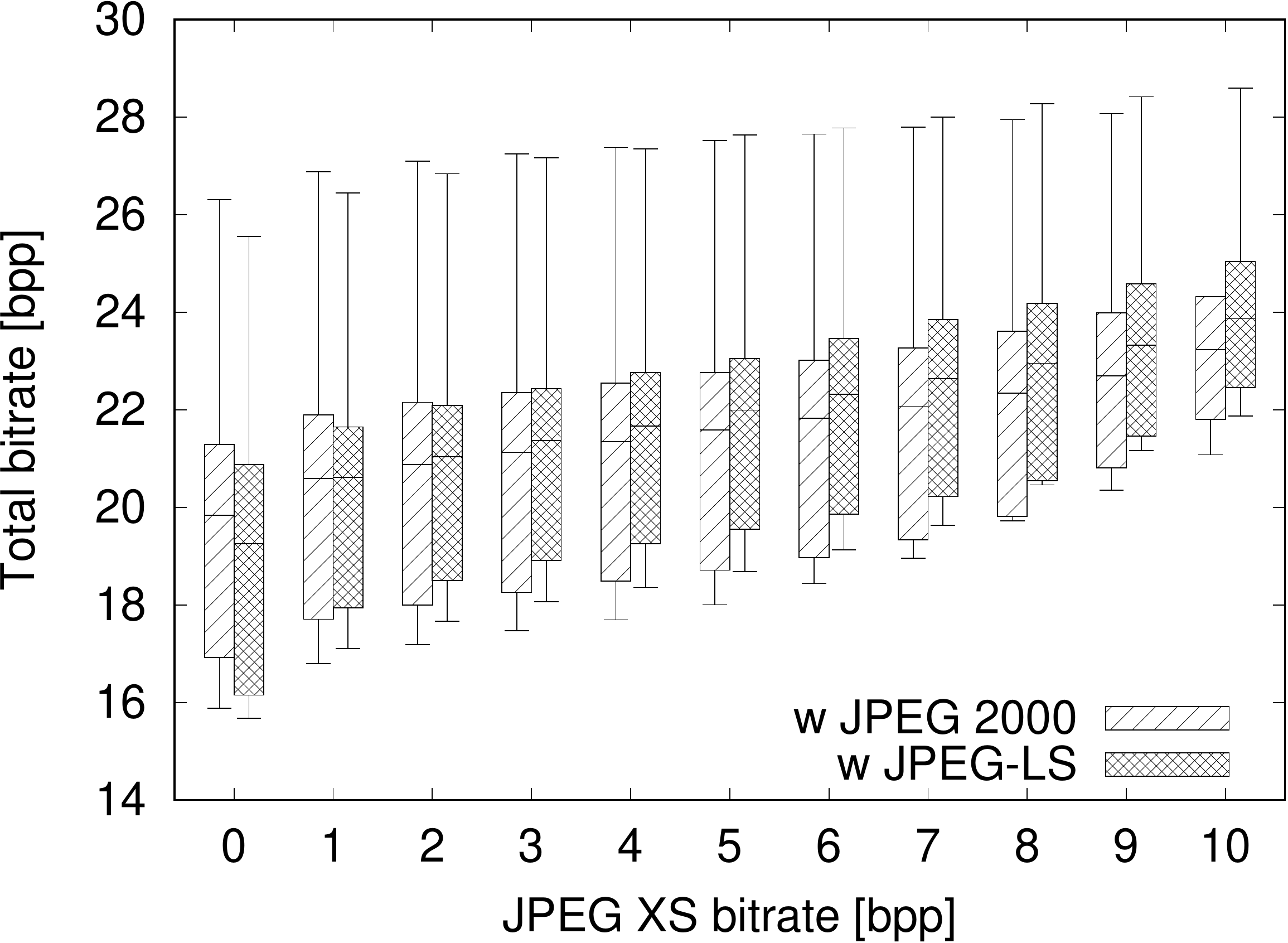} \\
  (b) 4K-images \\
 \end{tabular}
 \caption{Boxplot of total bitrates of the proposed coding with JPEG 2000 / JPEG-LS}
 \label{fig:LSvs2000}
\end{figure}

\wfigure{LSvs2000} shows the difference between total bitrate values of the two-layer coding with JPEG 2000 and those of with JPEG-LS.
From the figure, JPEG 2000 provided higher compression performance than with JPEG-LS.
In particular, when the bitrate value of JPEG XS was larger, the difference between them was larger.

\section{Conclusions}
We proposed a novel two-layer lossless coding method with backward compatibility to JPEG XS.
In the proposed coding, JPEG XS is used as the base layer, and the difference between an original image and the base layer image is encoded in the extension layer by using an arbitrary lossless encoder.
In the experiment, the proposed method was confirmed to achieve lossless coding under all conditions.
The difference between with JPEG 2000 and with JPEG-XS was also compared.


\bibliographystyle{IEEEbib}

\bibliography{library}

\end{document}